# Room-Temperature Charge Stability Modulated by Quantum Effects in a Nanoscale Silicon Island

S. J. Shin,[1] J. J. Lee,[1] H. J. Kang,[1] J. B. Choi,[1*] S.-R. Eric Yang,[2] Y. Takahashi,[3] and D. G. Hasko[4]

[1]Department of Physics & Research Institute for Nano Science & Technology, Chungbuk National University, Cheongju 361-763, South Korea.

[2]Department of Physics, Korea University, Seoul, 136-701, South Korea.

[3]The Graduate School of Information Science and Technology, Hokkaido University, Sapporo 060-0814, Japan.

[4]Cavendish Laboratory, Univ. of Cambridge, Cambridge CB3 0HE, United Kingdom.

* e-mail: jungchoi@chungbuk.ac.kr\

We report on transport measurement performed on a room-temperature-operating ultra-small Coulomb blockade devices with a silicon island of sub-5nm. The charge stability at 300K exhibits a substantial change in slopes and diagonal size of each successive Coulomb diamond, but remarkably its main feature persists even at low temperature down to 5.3K except for additional Coulomb peak splitting. This key feature of charge stability with additional fine structures of Coulomb peaks are successfully modeled by including the interplay between Coulomb interaction, valley splitting, and strong quantum confinement, which leads to several low-energy many-body excited states for each dot occupancy. These excited states become enhanced in the sub-5nm ultra-small scale and persist even at 300K in the form of cluster, leading to the substantial modulation of charge stability.

Keywords: single-electron transport, nanoscale silicon dot, Coulomb blockade, room-temperature charge stability, quantum effects

An ultra-small Coulomb blockade device can be regarded as a mesoscopic artificial atom system. Tunneling through it can provide a rich experimental environment for studying quantum transport phenomena.[1] Previously, these quantum effects have been investigated using relatively large devices at low temperatures, where they give rise to additional fine structures on the Coulomb oscillations.[2-9] However, as temperature increases up to 300K such fine structures observed at low temperature normally vanish together with Coulomb peaks themselves because of the weak Coulomb charging energy due to the relatively large dot size. However, as the dot size is reduced below 5nm, the very small number of electrons on the dot is expected to ensure that electron-electron interactions with Pauli spin exclusion strongly influence the electron transport characteristics. Here, we report on an extensive transport measurement performed on a room-temperature-operating Coulomb blockade device with an ultra-small silicon island of sub-5nm size. Transport data exhibits a striking feature of that the main room-temperature characteristics of the Coulomb peaks persist even at ultra-low temperature down to 5.3K. Substantial change in slopes and diagonal size of the room-temperature Coulomb diamond and bias-dependent peak splitting must reflect low energy many-body excited states associated with total spin for each dot occupancy N. This quantum effects become enhanced in our ultra-small Coulomb island and persist even at room-temperature, leading to the substantial modulation of the charge stability for finite bias window.

   The Coulomb blockade device used in transport measurement has been fabricated by scaling a state-of-the-art finFET structure[10] down to an ultimate form, by using deep-trench and subsequent oxidation-induced strain, which can be used to form a single electron transistor (SET) with a Coulomb island of sub_5nm size.[11] Figure 1(a) shows a SEM image of the SET device whose active channel is detailed as a schematic 3-D layout (Fig. 1(b). Note how the top-Si nanowire, exposed by the nano-gap between the source and the drain, is further etched down to 30 nm in depth by dry etching and gate oxidation. This key process, different from the conventional nanoscale finFET, enables a Coulomb island to be formed with nearly identical tunnel barriers in a self-aligned manner. Moreover, by wrapping a fin-gate almost completely around the Coulomb island, good control of the local electron potential is maintained. Figures 1(b) also shows a TEM cross-sectional images of the etched top-Si nanowire along the channel, exhibiting the island diameter of ~2-nm. Good control over the island size was achieved through the oxidation process.[11] Figure 1(c) shows the drain current measured at 300K as a function of the fin-gate voltage $V_G$ for SET, which is compared with that of the conventional finFET that was fabricated by similar process in the same wafer, but without the deep-trench process on the silicon wire channel. Note that the 1st Coulomb peak of the SET appears just above the threshold of the finFET that occur at Vg~2.5V for drain bias voltages up to 50mV. The onset of the Coulomb oscillations was not detected below the threshold voltage, indicating the 1st peak to be associated with the first electron tunneling.

Figure 2 shows temperature dependence of the I-Vg characteristics of the SET measured for various temperatures down to 5.3K for a bias 50mV. As seen in Fig. 2(a), the main feature of the Coulomb oscillations of 300K persists even at low temperature down to 5.3K, except for additional splitting observed in each Coulomb peak. We point out that this temperature-dependent feature is quite new and remarkable because the peak splitting so far observed at low-temperatures have been reported to vanish together with Coulomb peaks themselves with increasing temperature. This strongly indicates another evidence for that the Coulomb island of our device is a quite small well-defined single dot (of sub-5nm size), providing its charging energy to be large enough to get over the room-temperature thermal energy. If the main four Coulomb peaks were due to some multiple defects at Si/SiO2 interface, some of peaks should be randomly created or disappear depending on their thermal activation energies as temperature changes. Note also that the magnitude of the Coulomb peak decreases for low temperatures, as seen in Fig. 2(e) which shows the temperature-dependent magnitude of the 3rd Coulomb peak for each bias up to 50mV. This can be attributed to the possible decrease of the carrier concentration (or, carrier freeze-out) in the S/D wires as temperature decreases down to 5.3K.

Charge stability plots (displayed for down to 100K) are seen in Fig. 2b, 2c, and 2d, respectively, where successive Coulomb diamonds are clearly seen. Each diamond corresponds to a stable charge configuration state with fixed electron occupancy N. Coulomb peak splitting are seen for diamonds of N=3 and 4 in the charge stability even at 100K. Note that Coulomb diamonds for each N are very symmetric with respect to the positive and negative drain biases, strongly indicating that a single ultra-small Coulomb island is formed at middle point of the channel and that its tunnel barriers with source and drain are nearly identical.[12] This rules out a possibility of that the observed Coulomb oscillations may be related to the possible dopant or defect which must be randomly formed. The charge stability data also exhibit that as the gate voltage is made less positive, the slope of each Coulomb diamond steeply increases, and the Coulomb diamond (for $V_G$<3V) does not close. This feature is consistent with the lack of any Coulomb peaks below the threshold (Fig. 1(c)), indicating that the island is unpopulated by electrons for $V_G$≤3V. To more convince the assignment of the dot occupancy, a charge sensing device by means of a separated circuit (such as an additional SET, or quantum point contact) [13,14] must be installed next to the Coulomb island. In this case, however, an additional sensing bridge gate should be designed to be located very close to the dot to maximize mutual charge coupling. This, without doubt, yields a substantial increase of the total capacitance of the Coulomb island, leading to the SET operating only at ultra-low temperatures. Any kinds of room- temperature features will vanish. This is why most of experiments based on these devices including charge sensor have shown Coulomb oscillation behavior only at dilution refrigerator temperature of 10-100mK.[13-16] We, therefore, addressed the assignment of the dot occupancy by somewhat indirect ways (mentioned above) without using additional charge sensor.

It is noted that substantial change in slopes and diagonal size of each successive diamond is observed, implying that the charging energy is not constant over the gate voltage range studied. This behavior could be accounted for by strong interplay of the Coulomb interaction and additional quantum effects associated with very low electron number on the island. The size of the island and its Coulomb charging energy can be roughly estimated using the 1$^{st}$ diamond associated with the lowest dot occupancy N=1 that is determined mainly by the Coulomb charging energy. Values for the gate and junction capacitances can be directly obtained from the Coulomb peak spacing $\Delta V_G$ and the slopes of the 1$^{st}$ diamond.[1,11] This yields the total capacitance $C_\Sigma$ ~0.42aF, which corresponds to a 1.94nm diameter spherical silicon dot, in good agreement with the TEM image in Fig. 1(b). The charging energy of a dot of this size is thus $e^2/C_\Sigma$ ~0.38eV, which is more than one order magnitude larger than the thermal energy at 300K.

The fine structure with decreasing temperature must reflect low energy excited levels associated with each dot occupancy N, and can be explored more in detail with increasing bias window. Figure 3a and 3b are charge stability data at 5.3K. They illustrate the fine structure of the bias dependence of the Coulomb oscillations, showing typical behaviour of increasing splitting with bias window. For more clarity, we present Fig. 3c and 3d, reproducing Id-Vg for some specific bias voltages in the charge stability data. As seen in Fig. 3a with Fig. 3c, when bias voltage increases up to 100mV, the 1$^{st}$ peak starts to split into two sub-peaks and persists even at high bias, while the 2$^{nd}$ peak splits into 4 sub-peaks. Note that for the 1$^{st}$ main peak, the valley between two sub-peaks is raised up with bias voltage, indicating the increase in tunnelling current as bias widow becomes wide. This is not due to a peak broadening effect because the heating energy by increasing bias is only about ~0.05nW, negligible compared to the thermal energy of 5.3K. Similarly, Fig. 3b with Fig. 3d illustrate that the number of splitting of the 3$^{rd}$ Coulomb peaks rapidly increase from three to more than 12, while that of the 4$^{th}$ peak increases from two to more than 8. This strong bias dependence of peak splitting demonstrates the evident transition of the transport behaviour of our device from linear to non-linear transport regime where single-electron tunnelling can be made through many excited levels lying within the bias window.[17,18] It is thus important to know the low energy level spectrum associated with each dot occupancy N to explain the observed fine structure in each main Coulomb peak.

The Coulomb channel in our device fabricated on (100) Si-2D system is surrounded by SiO2 insulator and is along <110> direction that is parallel to the notch orientation axis of our SOI wafer. In such a wire valley splitting lifts the twofold and fourfold degeneracies into two, and the energy levels of Γ valleys are lower than those of off-Γ valleys.[19,20] In Si valley splitting is a main source of spin decoherence. It has been observed even in zero external electric field in strongly confined nanostructures and could be further enhanced by the application of bias, strain, or magnetic field.[21,22] In the presence of a confinement potential along the wire the energy levels of valleys are quantized, as shown schematically in the inset (a) of Fig.4. With the inclusion of the valley splitting the four lowest energy levels of the dot are $\varepsilon_1 = E_{v2,1}$, $\varepsilon_2 = E_{v2,1} + \Delta$, $\varepsilon_3 = E$ and $\varepsilon_4 = E + \Delta'$, where $E_{v2,1}$ is the lowest energy quantized with valley splitting $\Delta$, and E is the second lowest energy quantized with valley splitting $\Delta'$. While $E_{v2,1}$ and $E_{v2,1}+\Delta$ originate from $\Gamma$ valley the energy E may originate either from $\Gamma$ or off-$\Gamma$ valleys.[17,18]

Based on the above information on single electron levels we model the many-body Hamiltonian[23] of the dot by

$$H = \sum_i \varepsilon_i n_i + \sum_{i<j} V_{ij} n_i n_j + \sum_i U_i n_{i\uparrow} n_{i\downarrow} - \sum_{i<j} J_{ij} (\vec{S}_i \cdot \vec{S}_j + \tfrac{1}{4} n_i n_j) . \quad (1)$$

Here the label i=1, 2, 3, 4 denotes the four single electron states. For each single electron level i we define, respectively, the quantities $\vec{S}_i$, $n_{i\sigma}$, $n_i$, and $U_i$ as the spin operator, number operator of electrons with spin $\sigma$, number operator of occupied electrons, and the intra-level Coulomb repulsion. The Coulomb repulsion (exchange) energy between an electron in the i'th and an electron in j'th levels is $V_{ij}$ ($J_{ij}$). Each many-body eigenstate can be represented by a ket state $|\{n_i\}, S, S_z\rangle$. Level occupation numbers $\{n_i\}$ and the total spin quantum number S of some of the lowest energy many-body states are analysed and displayed in Fig. 5. (note that since the Hamiltonian is spin rotationally invariant eigenvalues are independent of the z-component total spin $S_z$). In this classification of eigenstates it is useful to exploit the fact that when the i'th level is doubly occupied its spin state is necessary a singlet state with $\vec{S}_i=0$. In addition, according to the quantum rules of spin addition, the total spin state S=1/2 of 3 electrons can be constructed by adding spin 0 and 1/2 or by substracting spin 1/2 from 1. This implies that, when adding 3 electron spins, there are two different spin wavefunctions for the same S=1/2 and $\{n_i\}$. Due to many-body exchange interactions states with the same $\{n_i\}$ but with different S do not have the same energy. For example, for N=2 the S=0 singlet and S=1 triplet states are split, see Fig. 5(b). From our observed data, Fig. 3(c), the magnitude of the exchange interaction is estimated to be about 0.01eV. Using these rules we find, respectively, 2, 4, 12, and 8 number of lowest energy states for N=1, 2, 3, and 4 that can be formed from the energies $\varepsilon_1, \varepsilon_2, \varepsilon_3$, and $\varepsilon_4$, see Fig. 5(a)-(d). We stress that these numbers are independent of model parameters. For a given N, estimate of energies shows that there is an energy gap to the next excited states from the group of lowest energy states mentioned above.

For N=1 and 2 the theoretical number of lowest energy levels agree with experimentally observed peak values of 2 and 4. According to our model for N=2 there are 3 singlet and 1 triplet levels, see Fig. 5(b). The three singlet states have all different energies because their occupation number configurations $\{n_i\}$ are different. For N=3 we displayed 12 lowest energy states in Fig. 5(c). In near agreement with this value the observed number low energy excited states is about 12-14 (when small noise-like peaks are included there are 14 peaks). For N=4 some of the lowest excited energy states have the same $\{n_i\}$ but different spin values S=0 and 1, see Fig. 5(d). Their energies are again different because these states have different spin wavefunctions and, therefore, have different exchange energies. For N=4 the observed number of excited states is between 8 and 10, which is close to predicted value of 8, see Fig.5(d). Since there may be other single electron energy levels close to $\varepsilon_3$ and $\varepsilon_4$ different many-body excited energies may be present near those 12 and 8 states shown in Figs. 5(c) and 5(d). Moreover, the effect of quantum fluctuations of occupation numbers, which is absent in Hartree-Fock approximations, may give rise to additional excited states[24]. These factors suggest a possible explanation for why more than 12 and 8 peaks may have been observed for N=3 and 4, respectively, when small noise-like peaks are included.

The Coulomb interaction energy between two electrons in the level 1 is $U_1 \approx \dfrac{e^2}{\varepsilon R} \approx 0.38\text{eV}$, where $\varepsilon$ and R are the dielectric constant and the radius of the dot. The energy separation between different Coulomb peaks can be fitted by choosing two parameters $V_{13}$ and $\varepsilon_3 - \varepsilon_2$ judiciously: $V_{13} \approx 0.3\text{eV}$ is the Coulomb interaction energy between two electrons in the levels 1 and 3 and $\varepsilon_3 - \varepsilon_1 \approx 0.3\text{eV}$ is the energy separation between them. The fit value of $V_{13}$ is reasonable because the inter-level Coulomb interaction is comparable to the intra-level $U_i$, but must be smaller than it. The fit value of the quantum confinement of $\varepsilon_3 - \varepsilon_1$ is also in the range of expected value for our dot size of 2nm by theoretical calculations.[12,13] Using these values we find that the addition charging energies of N=1 $\rightarrow$ 2, 2 $\rightarrow$ 3, and 3 $\rightarrow$ 4 are approximately $U_1 \approx 0.38\text{eV}$, $2V_{13} + \varepsilon_3 - \varepsilon_2 \approx 0.9\text{eV}$, and $V_{13} + 2(U_1 - V_{13}) \approx 0.46\text{eV}$, which are illustrated in Fig. 4. Using the energy coupling factor, defined by $\alpha_G=C_G/C_\Sigma \sim 0.22$, the corresponding observed values 0.37eV, 0.84eV, and 0.50eV (see the inset (b) in Fig. 4) are in the same range. For N=1 the observed Coulomb oscillations exhibit splitting of the 1$^{st}$ peak is due to the valley splitting, see Fig. 5(a). The measured value of valley splitting $\Delta$ is 16meV, whose order of magnitude is consistent with recently reported theoretical values of sub-3nm Si nanostructures[19,20]. Note that its value is much smaller than Coulomb charging and quantum confinement energies. These approximate agreements between experimental and theoretical values suggest that our model can account consistently for several features of excited states in the ultra-small Si dot formed along <110> direction.

In summary, we report on an extensive transport measurements performed on the room-temperature-operating ultra-small silicon SET devices with a Coulomb island of sub-5nm size. The room-temperature feature of I-Vg persists even at low temperature down to 5.3K, where additional fine structures of Coulomb peaks appear. The unusual energy separation between Coulomb diamonds and the fine splitting of each Coulomb peak are accounted for by including quantum many-body interactions, leading to the substantial modulation of Coulomb diamonds at 300K. It further supports the reliability in our CMOS-compatible implementation of the ultra-small SET operating at room-temperature.

**Acknowledgement:** This work was supported by the National Research Foundation through the Frontier 21 National Program for Tera-level NanoDevices, Global Partnership Research Program with University of Cambridge, and in part by a research grant of the Chungbuk National University in 2009.

**Figure Captions:**

**Figure 1:** (**a**) SEM image of the Coulomb blockade SET device. (**b**) Schematic 3-D layout of the active channel area of the device and a cross-sectional TEM images along the channel, showing Coulomb island size of ~2nm. (**c**) Comparison of the I-Vg characteristics of SET with those of the conventional nano finFET for drain bias up to 50mV at 300K.

**Figure 2:** (**a**) Temperature dependence of the I-Vg characteristics of the SET measured for various temperatures down to 5.3K for a bias Vd=50mV. Note that the main feature of 300K persists even at low temperature down to 5.3K, but a striking temperature-dependent splitting is observed in each Coulomb peak. (**b**), (**c**) & (**d**) Charge stability plot for temperatures of 300K, 200K and 100K, respectively. Each Coulomb diamond corresponds to a stable charge configuration state with fixed electron occupancy N. Peak splitting are clearly seen for diamonds of N=3 and 4 even at 100K. (**e**) Temperature-dependent magnitude of the 3$^{rd}$ Coulomb peak for each bias up to 50mV.

**Figure 3:** Charge stability plot at 5.3K and specific bias dependence of each main Coulomb peak; (**a**) & (**b**) Charge stability plot at 5.3K, showing typical behaviour of increasing splitting with bias window. (**c**) & (**d**) I-Vg characteristics for some specific bias voltages, which are reproduced from the charge stability data. Strong bias dependences of peak splitting are clearly seen, which can be accounted for by the non-linear transport made through many excited levels associated with each dot occupancy N.

**Figure 4:** Addition charging energies of N=1→2, 2→3, and 3→4, estimated from the many-body Hamiltonian. The low energy level spectrum associated with each dot occupancy N are illustrated. Inset (**a**) illustrates a confinement potential along the wire where the energy levels of valleys are quantized. **T**he calculated addition charging energies, approximately $U_1 \approx 0.38$eV, $2V_{13} + \varepsilon_3 - \varepsilon_2 \approx 0.9$eV, and $V_{13}+2(U_1 - V_{13})$

$\approx 0.46$eV (for N=1→2, 2→3, and 3→4, respectively) are denoted by arrows in inset **(b),** which are in the same range as those of the charge stability data observed at 300K.

**Figure 5:** Electronic occupation configurations illustrating the number of lowest energy states for **(a)** N=1, **(b)** N=2, **(c)** N= 3, and **(d)** N=4, respectively. Note that due to many-body exchange interactions states with the same $\{n_i\}$ but with different S do not have the same energy.

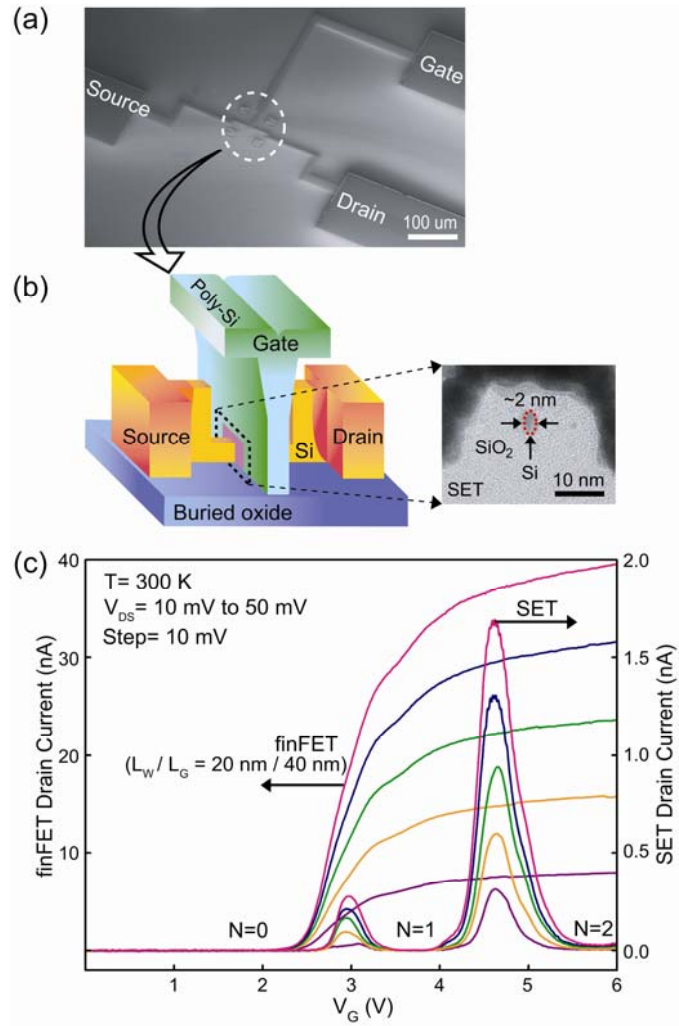

**Fig. 1_S. J. Shin *et al.***

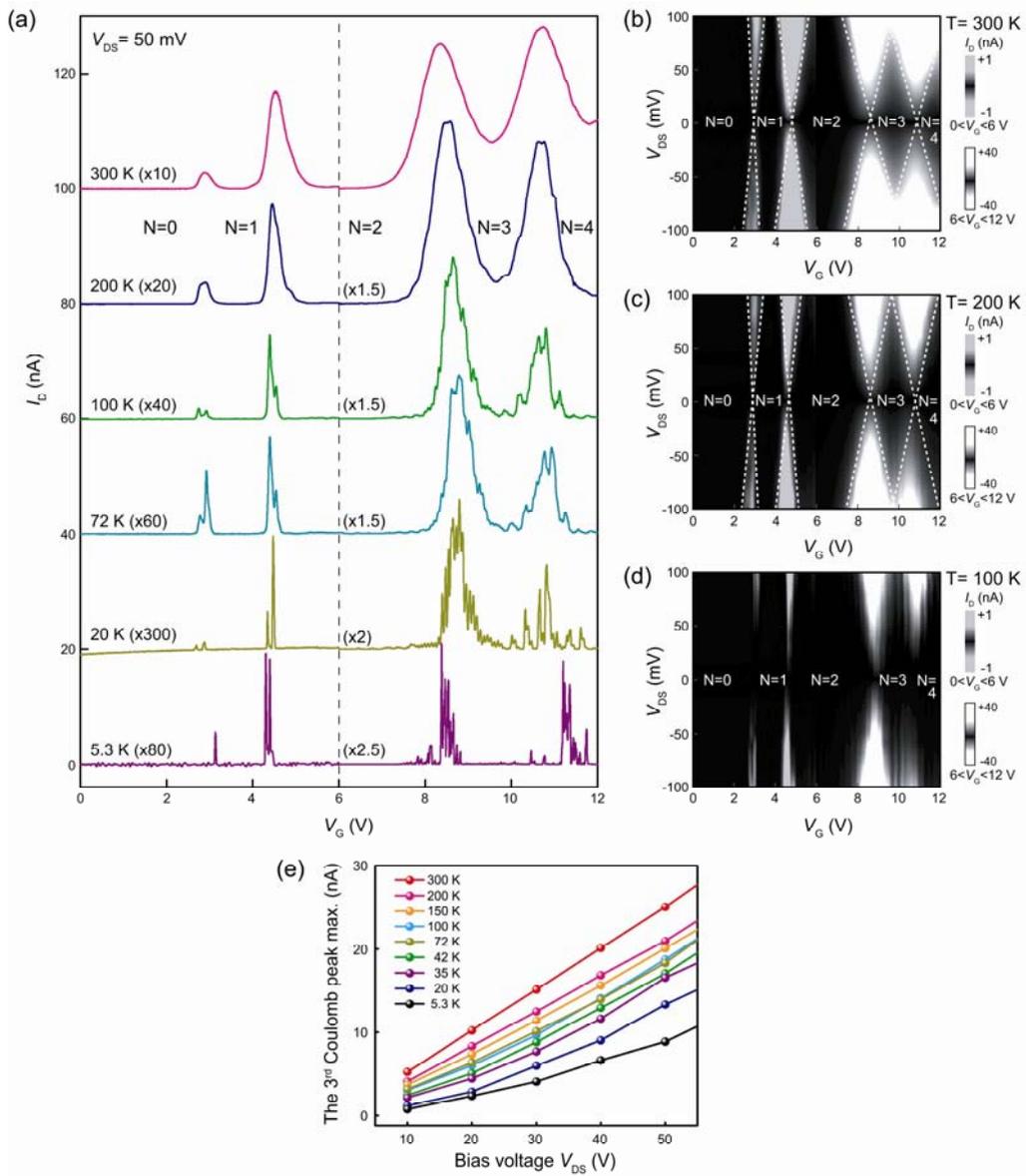

**Fig. 2_S. J. Shin *et al.***

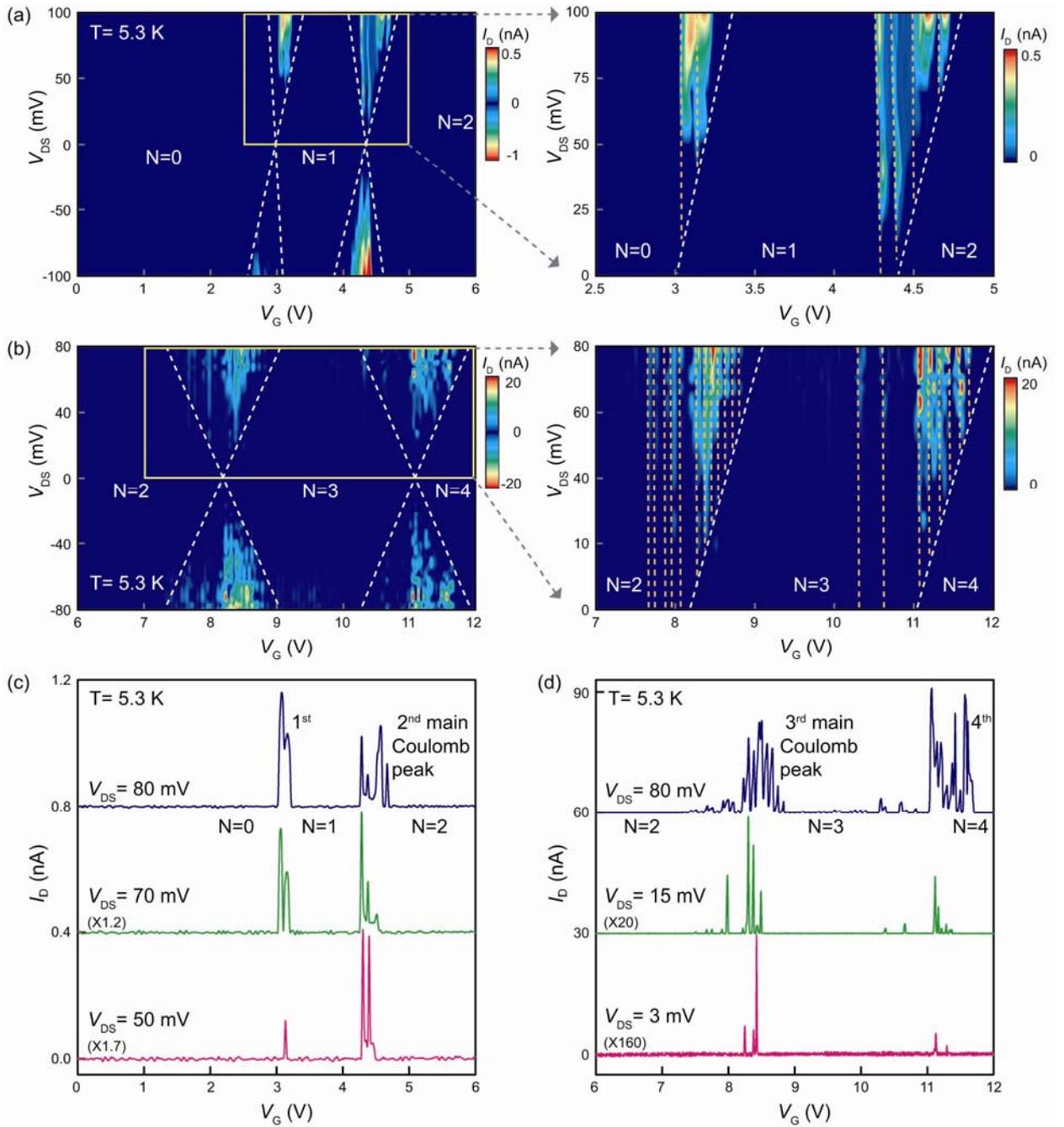

**Fig. 3_S. J. Shin *et al.***

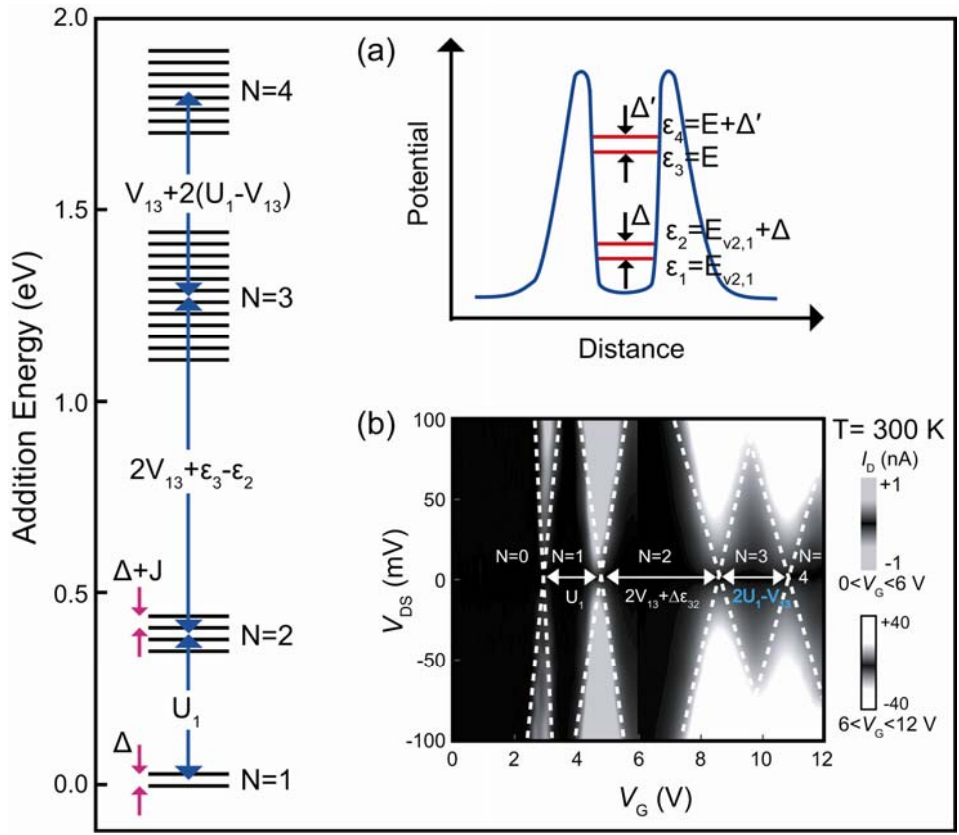

**Fig. 4_S. J. Shin *et al.***

**Fig. 5_S. J. Shin *et al.***